# Optical detection of bacterial cells on stainless-steel surface with a low-magnification light microscope

**Authors:** Yuzhen Zhang[a], Zili Gao[a], and Lili He[*]

**Affiliations:**

[a]Department of Food Science, University of Massachusetts, Amherst, MA 01003, USA.

[*] Co-corresponding authors:

Dr. Lili He, E-mail: lilihe@foodsci.umass.edu; Fax: +1 413 545 1262; Tel: +1 413 545 5847

**Abstract**

A Rapid and cost-effective method for detecting bacterial cells on surfaces is critical to protect public health from various aspects, including food safety, clinical hygiene, and pharmacy quality. Herein, we first established an optical detection method based on a gold chip coating with 3-mercaptophenylboronic acid (3-MPBA) to capture bacterial cells, which allows for the detection and quantification of bacterial cells with a standard light microscope under low-magnification (10 ×) objective lens. Then, integrating the developed optical detection method with swab sampling to achieve to detect bacterial cells loading on stainless-steel surfaces. Using *Salmonella enterica* (SE1045) and *Escherichia coli* as model bacterial cells, we achieved a capture efficiency of up to $76.0 \pm 2.0$ % for SE1045 cells and $81.1 \pm 3.3$ % for *E. coli* cells at $10^3$ CFU/mL upon the optimized conditions. Our assay showed good linear relationship between the concentrations of bacterial cells with the cell counting in images with the limit of detection (LOD) of $10^3$ CFU/mL for both SE1045 and *E. coli* cells. A further increase in sensitivity in detecting *E. coli* cells was achieved through a heat treatment, enabling the LOD to be pushed as low as $10^2$ CFU/mL. Furthermore, successful application was observed in assessing bacterial contamination on stainless-steel surface following integrating with swab collection, achieving a recovery rate of approximately 70 % suggests future prospects for evaluating the cleanliness of surfaces. The entire process was completed within around 2 hours, with a cost of merely $ 2 per sample. Given a standard light microscope cost around $ 250, our developed method has shown great potential in practical industrial application for bacterial contamination control on surfaces in low-resource settings.

**Key words**: low-magnification microscopy, 3-mercaptophenylboronic acid, bacterial cells, stainless-steel surface, swab sampling

## 1. Introduction

The assessment of bacterial cells on surfaces is crucial across diverse sectors, such as food processing industries (Cosby et al., 2008; De Filippis et al., 2021; Sibanyoni and Tabit, 2019), healthcare environments (Dancer, 2009, 2004) and pharmaceutical factories (Tršan et al., 2019) for some reasons. For example, food contact surfaces play an essential role in food safety since bacterial cross-contamination can take place during food producing or manipulating (Guillermo Ríos-Castillo et al., 2021; Lim et al., 2021). Effective evaluations of bacteria on food contact surfaces can prevent the spoilage and extend the shelf life of food products, mitigate the spread of foodborne illness as well (Cosby et al., 2008; Moore et al., 2007; Sibanyoni and Tabit, 2019). Similarity, detection of bacteria on surfaces within healthcare settings contributes to infection prevention and control to some extent as it has been reported that the reduction in the frequency of hospital-acquired infections following effective disinfection measures for surface contamination (Axel Kramer and Ojan Assadian, 2014). In pharmaceutical manufactures, the monitoring of bacteria on surfaces is imperative to uphold the quality of pharmaceutical products, thereby minimizing health hazard to the patients (Stephen P. Denyer and Rosamund M. Baird, 2006). Regular assessment of bacteria on surfaces not only guide and verify the effectiveness of disinfection and sterilization measures in controlling the growth and spread of bacteria on those surfaces, but also make sure public health by improving food/clinical safety and products quality.

The current methods for assessing bacteria on surfaces have major drawbacks. In practical settings, two main methods are commonly applied: aerobic plate count (APC) and adenosine triphosphate (ATP) bioluminescence swab kit. Even though APC is known as standard method to indicate bacterial population in samples, it typically demands a considerable timeframe of 2 to 3 days to get the result which is time-consuming. The drawback not only influences food manufacturing, where considerations of limited shelf life and production efficiency of food products, but also public surroundings which accommodate a high daily influx of visitors (Ferone et al., 2020; Ma et al., 2023). Another limitation is the difficulty to recover growth of injured and viable but non-culturable (VBNC) bacterial cells, leading to the potential false-negative results (Everis, 2001; Li et al., 2014). On the other hand, ATP swab kit offered simple and rapid measurement based on correlation of the ATP concentration in objects with relative light units (RTU) (Omidbakhsh et al., 2014). However, several limitations were found during practical applications. One initial drawback is ATP meters are costly even though swab kits are affordable. Additionally, it is incapable to detect actual quantity of bacterial cells with the interference of other organic matters releasing ATP as well, like allergic protein and residual leafy green (Bakke, 2022; Lane et al., 2020). Some researchers also reported that the testing results can be affected by some factors: the incomplete bacterial lysis, disinfectants and cell viability (Bakke, 2022; Danielle E Turner et al., 2010; Lappalainen et al., 2000). Furthermore, the absence of standard correlations between bacterial concentrations and RTU among ATP swab kits from different manufacturers results in no specific standards or regulatory limits on RTU to define contamination levels (Andreas van Arkel et al., 2021). Lacking a rapid, effective and cost-effective method to detect bacteria on surfaces will hamper the identify and response to the potential sources of contamination in a timely manner, and cause large loss due to products recall. Thus, developing a rapid, economical and effective method is essential for ensuring environmental safety and protecting public from the risk of bacterial contamination in healthcare, food and pharmacy.

The objective of this study is to develop rapid and cost-effective method to detect bacterial cells on surfaces. The detecting strategy was based on 3-mercaptophenylboronic acid (3-MPBA) coated gold slides in our prior study which can capture bacterial cells through the formation of covalent bond between boronic acid group of 3-MPBA and 1,2 and 1,3 *cis* diols of glycans (polysaccharides and glycoproteins) on the outer surface of bacteria (Hickey and He, 2020; Pearson et al., 2017; Rao et al., 2016). Notably, it has been shown that 3-MPBA was able to capture bacterial cells nonspecifically regardless of Gram-negative or positive, which allows 3-MPBA coated gold chip to achieve the same detection purpose as APC and ATP for total bacterial counts (Tra and Dube, 2014). In addition, taking advantage of the hydrophobicity of 3-MPBA coated on gold chip to enhance the visualization of bacterial cells, the single bacterial cell can be observed under 20 × objective lens instead of typically using medium magnifications (50 - 100 ×) (Pearson et al., 2017). The usage of lower magnification objective lens is significant as it promises the method both affordable and easy to be operated by facilities staff.

In this study, we aimed to explore the utilization of a low-cost microscope with 10 × objective lens ($ 250) in the method development. *Salmonella typhimurium* and *Escherichia coli* were chosen as model bacterial cells for method optimization and evaluation. In the last, the application of the developed optical detection method in evaluating the level of bacterial cells loading on stainless-steel surface by coupling with swab sampling was also investigated. To the best of our knowledge, this is the first study that explored the feasibility of using a low magnification (10 × objective lens) optical microscope for the detection and quantification of bacteria recovered from surface swab.

**2. Experimental**

## 2.1 Preparation of bacteria

*Salmonella enterica* subsp. *enterica* BAA1045 (SE 1045) and *Escherichia coli* OP50 (*E. coli* OP50) were used in this study. The frozen cultures of SE 1045 and *E. coli* OP50 were revived by inoculating them onto tryptic soy agar (TSA) (Becton, Dickinson and Company, Difco, Franklin Lakes, NJ) and subsequently incubating them at 37 °C for overnight. A single colony was transferred into 10 mL of tryptone soy broth (TSB) (Becton, Dickinson and Company, Difco, Franklin Lakes, NJ) under 37 °C incubation until stationary growth phase was reached (approximately 16 hours). Bacterial cells were collected by centrifuging at 9000 × g for 5 minutes, followed by careful removal of the supernatant. Then, the cells re-suspended in 1 mL of double distilled autoclaved water and subjected to another round of centrifugation. This washing step repeated twice to eliminate the residue of TSB. After the final washing step, the supernatant was taken away and 1 mL of 50 mM ammonia bicarbonate (Fisher Scientific, Fair Lawn, NJ) was added to resuspend bacterial cells (approximately $10^9$ CFU/mL). A serial dilution of bacterial cells was implemented to achieve the desired concentrations using 50 mM ammonia bicarbonate. The dilutions were adjusted to a pH of 8.4 with 100 mM NaOH solution in a ratio of 1:10.

For the preparation of heat-treated samples, 1 mL of bacterial cells cultured for approximately 16 hours at 37 °C was subjected to thermal treatment. Exposing the sample to a temperature of 72 °C for 10 minutes in dry bath incubator (Fisher Scientific, Fair Lawn, NJ). The actual temperature was monitored with a thermometer introduced in another tube loading 1 mL bacterial cells inside the incubator. Once the temperature reached up to 72 °C, starting to count time. After the heat treatment, bacterial cells were washed and diluted as explained above.

## 2.2 Preparation of 3-MPBA coated gold chip

The protocol of coating 3-MPBA on a gold chip was adapted from Pearson et al.'s one (Pearson et al., 2017). Briefly, cutting the gold slide that was washed with ethanol into rectangular pieces (0.5 × 0.5 cm) with a glass knife. Then, each gold chip put into each well of 96-well plate and washed with ethanol twice. Removing the ethanol and waiting for 5 minutes to let the gold chip dry. 200 µL of 40 mM 3-MPBA (TCI America, Portland, OR) that dissolved in ethanol was pipetted into each well and make to interact on a shaker (speed = 20 rpm) (Fisher Scientific, Fair Lawn, NJ) for overnight (approximately 17 hours). On the subsequent day, the 3-MPBA solution was withdrawn, and 200 µL of distilled water was added to wash chips for 20 s before the surface of gold chip was dried. The washing step repeated once again. After the second washing, the supernatant was pipetted out and 200 µL of distilled water was added into each well to stabilize the chip until use.

## 2.3 Optical detection protocol

Once the solution of bacteria was prepared well, the distilled water in each well that contained a 3-MPBA coated gold chip was removed, followed by adding 50 µL of bacterial solution into the respective well. After incubation period of 1 hour of 3-MPBA coated gold chip with bacterial solution, the supernatant was taken away and washed twice with distilled water. After completing the last washing step, the water was pipetted out and let samples air dry for 5 minutes. Transferring each sample from each well to glass slide with tweezers carefully. The Olympus optical light microscope (Olympus Optical, Tokyo, Japan) was used to collect images with a 20 × objective lens. Another standard light microscope (AmScope, Irvine, CA) was also employed to take images using a 10 × objective lens.

**2.4 Data collection and analysis**

Placing the glass slide on the stage of optical microscope. Slowly rotating the coarse or fine adjustment knob until obtained a clear image. Each bacterial cell presented as "black dot" under the view of optical microscope. Every sample was taken at least 3-4 images with different regions. The plugin microbeJ was installed in software ImageJ and used to count number of "black dot" on each image (Ducret et al., 2016). The principle of the plugin counting is to tell the interest particles apart from background by setting threshold manually on a grey image. All pixels with values above the threshold will be marked. Due to the difference of brightness on one image, the certain value of threshold is unable to recognize all objects from various brightness of the background. To overcome this, Adobe Photoshop 2022 was used to select region with uniform brightness, as well as converting the images into 8-bit grayscale before importing into ImageJ. Importing the processed images into ImageJ, adjusting threshold to select each "black dot". Several parameters of particles also were defined to exclude the noise particles on the image, such as round shape and a size range spanning from 3 to 60 pixel$^2$. In instances where several particles tightly contacted, the segmentation option was employed, which automatically separated them in accordance with the predetermined parameters of particles. As a result of counting, the edge of counted particles were marked with solid green line, as well as the summary window including the total counts.

**2.5 Measurement of zeta potential of bacteria**

The zeta potential of bacteria is not only influenced by the composition of the surface but also the nature of the surrounding medium such as ionic strength and pH (Soon et al., 2011). To minimize effects of medium, 0.5 mM potassium phosphate buffer solution (pH 7.4) was used to resuspend

$10^8$ CFU/mL bacterial cells. The suspension was placed in the clear disposable zeta cells (Malvern Panalytical, Westborough, MA) prior to zeta potential measurement. The Zetasizer Nano-ZS device (Malvern Instruments, Worcestershire, UK) was conducted to measure the direction and velocity of the bacterial cells movement when subjected to the applied electric field at room temperature (25 °C). Each of the measurements was carried out under identical conditions (n = 3). Based on the Helmholtz – Smoluchowski theory (Hiemenz, 1977), the instrument's software program converted the recordings into zeta-potential values.

**2.6 Collection of bacteria from stainless steel surface using swab**

Prior to drop the bacterial cells on the stainless steel (SS) 304 coupons (10 × 10 cm), the surface was washed by 70 % ethanol twice. After drying, 100 μL of $10^5$ CFU/mL SE1045 was distributed in small droplets evenly with a pipette (approximate 16 droplets per area). The Fisherbrand™ PurSwab foam swab (Fisher Scientific, Fair Lawn, NJ) was homogeneously pre-moistened by submerging the head into distilled water, and pressed against the tube to remove the excess liquid before swabbing. The moistened swab swept the surface in an overlapping slalom-like pattern twice over the area. The second swabbing perpendicular to the first sweeping direction. The collected bacterial cells released from swab into 400 μL of distilled water by vortex vigorous for 1 min. The swab pressed against the tube to squeeze all the liquid before taking out. The bacterial cells were pelleted by centrifugation at 9000 × g for 5 min, resuspend into 45 μL of 50 mM ammonia bicarbonate, and 5 μL of 100 mM NaOH was added to adjust the pH prior to incubate with 3-MPBA coated gold chip. Incubating recovered bacterial cells with 3-MPBA coated gold chip, and collecting images with 10 × objective lens, followed by section 2.3 and 2.4. The recovery rate of our method to detect bacterial cells loading on stainless-steel surface was calculated using equation:

$$\text{Recovery rate (\%)} = \frac{N_{\text{bacterial cell (recovered from surface)}} - N_{\text{negative control (recovered from surface)}}}{N_{\text{cultured bacterial cell}} - N_{\text{Negative control}}} \times 100$$

Where $N_{\text{bacterial cell (recovered from surface)}}$ is the number of bacterial cells counted in images of samples recovered from stainless-steel surface, $N_{\text{negative control (recovered from surface)}}$ is the number of particles counted in images of samples recovered from stainless-steel surface without any bacterial cells. $N_{\text{cultured bacterial cell}}$ is the number of bacterial cells counted in the images of samples where cultured bacterial cells were introduced to incubate with 3-MPBA coated gold chip, $N_{\text{Negative control}}$ is the number of particles counted in the images of samples where ammonia bicarbonate without bacterial cells was added to incubate with 3-MPBA coated gold chip.

## 3. Results and discussion

### 3.1 Optimization of the 3-MPBA based optical detection protocol

Figure 1 illustrated the depiction of the optical detection procedure for bacterial cells using a 3-MPBA coated gold chip, which mainly consists of: 1. coating 3-MPBA with gold chip; 2. washing away the excessive 3-MPBA residue from the gold surface, 3. incubating bacterial cells with the 3-MPBA coated gold chip within an ammonia bicarbonate solution, 4. further washing to remove crystals of ammonia bicarbonate and any potential clumped bacterial cells. To ensure adequate coverage of 3-MPBA on the gold chip and maximize the capability to capture the majority bacteria, a range of different concentrations of 3-MPBA were used to modify gold chip and then incubated with *E. coli* cells at $10^8$ CFU/mL. As depicted in Figure 2a, the number of bacterial cells counted from images (Figure S1) using ImageJ exhibited the incremental trends as the 3-MPBA concentrations increased from 10 to 40 mM. The bacterial cell count remained constant at higher concentrations, suggesting that 40 mM 3-MPBA was enough to covered the gold surface in the size of 0.5 × 0.5 cm as much as possible. For the gold chips of larger dimensions, the concentration of 3-MPBA should be increased accordingly to guarantee the optimal coverage of gold chip, following the ratio of 0.16 M per $cm^2$.

After coating 3-MPBA with gold chip overnight, a significant accumulation of excess 3-MPBA remained on the gold surface, leading to the formation of numerous crystals if no washing step was implemented (Figure S2a). Thus, washing step and conditions should be considered including washing solutions and times. Given the favorable solubility of 3-MPBA in ethanol, we first examined the washing performance of ethanol. The 3-MPBA crystals were removed effectively by ethanol even only washed once (Figure S2b). Double distilled water was investigated as well, which gave to the clean gold surface after two washing cycles (Figure S2c, d). However, when it comes to the capability of capturing bacterial cells, the 3-MPBA coated gold chip washed by ethanol captured roughly 2 times lower than the one washed by double distilled water (Figure 2b). The reason for the discrepancy might be a portion of the 3-MPBA coated onto the gold chip was washed away because it dissolved well in ethanol. Therefore, using double distilled water to wash twice can clean 3-MPBA coated gold chip effectively without compromising the chip's capture capability.

Upon completion of the washing process for the 3-MPBA coated gold chip, an ammonia bicarbonate suspension containing $10^8$ CFU/mL *E. coli* was added to interact with the 3-MPBA coating. To understand the optimized duration for bacterial cells interacting with 3-MPBA, the different timepoints were investigated. It was obvious that the number of "black dots" on the images with 60 min incubation was greater than the one with 30 min incubation (Figure S3), and no significant increase even after a longer time (Figure 2c), illustrating 60 min was the best time for 3-MPBA capture bacteria cells. It was worth to note that adding the bacterial suspension into each well before the surface of the 3-MPBA coated gold chip dried achieves to the uniform distribution of bacterial cells (Figure S4a). On the contrary. the dry surface contributed to the clumping of the bacterial cells since the 3-MPBA molecules might become tangled up and entangled with each other, causing the boronic groups distributed unevenly (Figure S4b).

To achieve to capture as much as bacterial cell on the 3-MPBA coated gold chip, the more detailed condition of Step 3 was optimized. We speculated that the shaking speed during incubation effected the binding efficiency as it determined the collision probability of the bacterial cell with 3-MPBA molecular. The lower shaking speed might potentially yield the higher binding efficiency because the bacterial cells were easy to be settled down to increase the probability of contact with 3-MPBA molecular. To test this, the different shaking speeds - 0, 20 and 100 rpm - were applied. As we expected, the captured percentage of bacterial cells reached up to 60.6 ± 9.1 % of without shaking, surpassing the results at both 20 and 100 rpm (Figure S5).

Figure 2d1 and 2d2 shown the images of $10^8$ CFU/mL *E. coli* captured by the 3-MPBA coated gold chip, followed by washing with water for 0 and 1 time, respectively. It is unable to obtain the single "black dot" on both of images due to the crystals of ammonia bicarbonate that still remained on the surface and mixed with bacterial cells. The clear single "black dot" can be observed until washing up to twice, and there is no notable difference with the number of bacterial cells regardless of washing twice or even more (Figure 2d3 and S6). It indicated the binding affinity of glycan of bacterial cells and boronic group was so stable and strong that almost unaffected by washing multiple times.

It has been reported that the visual enhancement of the captured bacteria cells caused by 3-MPBA made it possible to effectively observe and enumerate captured bacterial cells using a 20 × objective lens (Pearson et al., 2017). It was a significant improvement that address the challenge of visualizing bacteria cells under bright field condition with low-magnification lens. Our results not only substantiate the reproducibility of the optical detection with a 20 × objective lens but also illustrated the capability of screening bacterial cells even with a 10 × objective lens (Figure S6). Remarkably, the price of the standard optical microscope utilized in this study was around $ 250 which is much cheaper than the elaborate microscope equipped with advanced objective lens and specialized settings. The affordability factor is the great advantage and improvement to widespread adoption among the manufactures especially for low-resource settings.

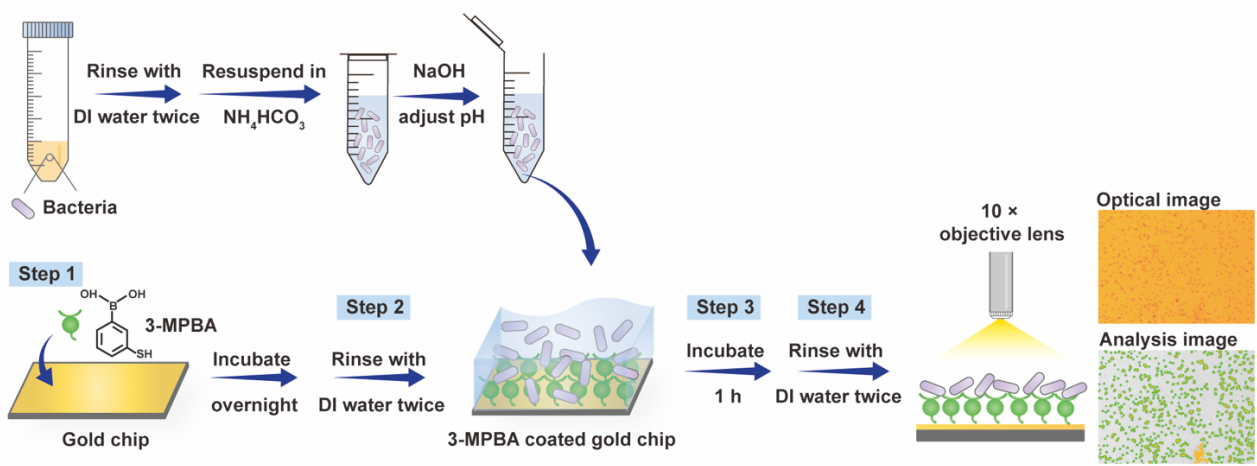

Figure 1. Scheme of principle of optical detection for bacterial cells based on the 3-MPBA as capturer.

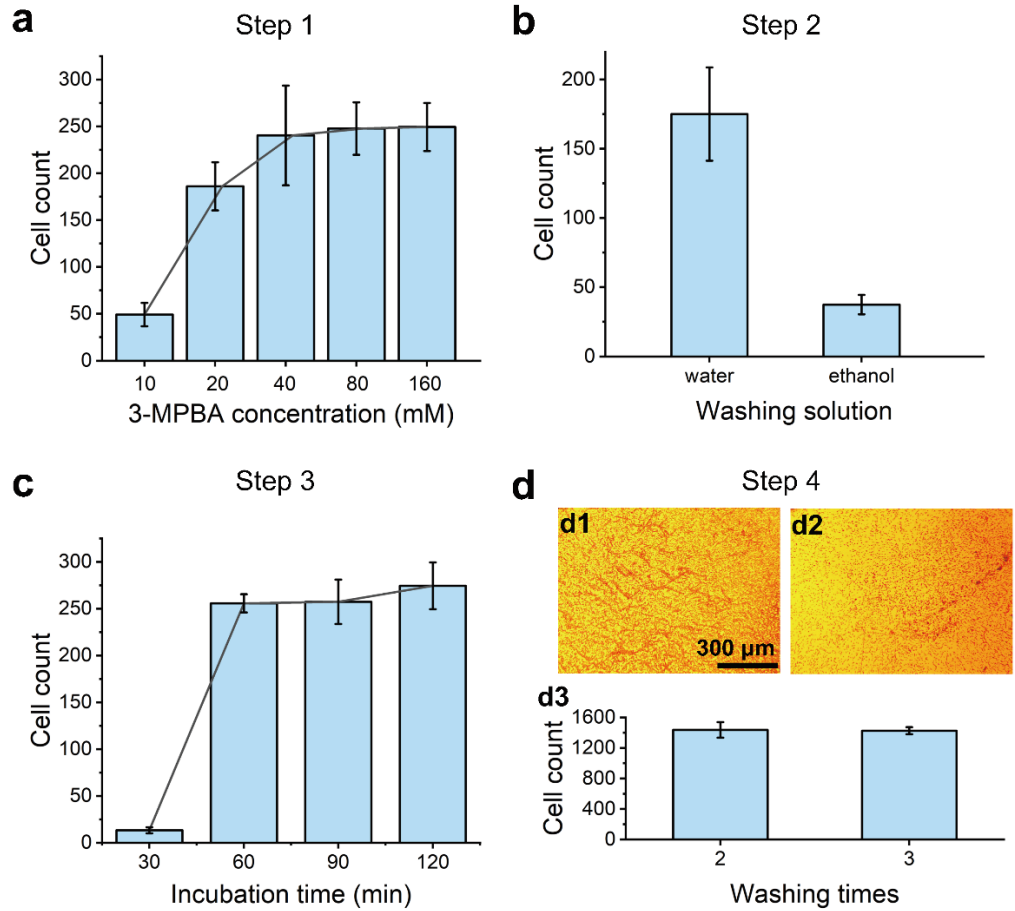

Figure 2. Optimization of the experimental conditions of the optical detection. Step 1, 2, 3, and 4 are described in Figure 1. (a). Cell counts of optical imaging results with different concentrations of 3-MPBA coating on gold chip. (b) Comparison of the cell counts of optical imaging results between using water and ethanol to wash gold chip after coating with 3-MPBA. (c) Cell counts of optical imaging results with different period of incubating bacterial cells with 3-MPBA coated gold chip. (d) Optical images (10 ×) and cell counts of optical imaging results with different washing times after incubation of bacterial cells with 3-MPBA coated gold chip: (d1) without wash,

(d2) washing once, (d3) comparison of the cell counts after washing twice and three times. Results were collected with SE1045 at $10^7$ CFU/mL. Scale bar represented 300 μm length. Data were represented as mean ± SD (n = 3).

## 3.2 Bacterial capture efficiency

Capture efficiency is a measurement of the proportion of the captured cells within the total population of bacterial cells, which is the vital feature of the assay to determine the sensitivity. As for evaluating the capture efficiency of the 3-MPBA coated gold chip, the number of bacterial cells before and after incubating with the 3-MPBA coated gold chip were detected by plate count colony method. Typically, the capture efficiency is expressed as the percentage of the difference in the number of bacterial cells before and after incubation relative to the initial number before incubation.

Figure 3a exhibited capture efficiencies of the gold chip with/without 3-MPBA coating for SE1045 at low, medium and high concentrations. The capture efficiencies of 3-MPBA coated gold chip were significantly ($p < 0.05$) higher than those of the uncoated gold chip overall, which was found with *E. coli* as well (Figure 3b). At low concentration ($10^3$ CFU/mL), the capture efficiency of 3-MPBA coated gold chip for SE1045 was 76 ± 2 %. It decreased with the increasing SE1045 concentrations in both chips. As with *E. coli*, the capture efficiencies of 3-MPBA gold chip were similar with that of SE1045 at low concentration. However, 23.1 ± 8.8 % of *E. coli* cells at $10^8$ CFU/mL and 44.7 ± 8.2 % at $10^5$ CFU/mL were captured by 3-MPBA coated gold chip, which were lower than the SE1045 cells. Theoretically, both of *E. coli* and SE1045 are gram-negative bacterium, sharing a large amount of the genetic material and a common structure of cell envelopes containing outer membrane layer and peptidoglycan layer (Hu et al., 2010). The existing researches have not illustrated the big variations of cell wall composition and lipid content between the two (Cozi et al., 1977; Heinrichs et al., 1998; Nakae and Nikaido, 1975). Based on our results, we hypothesize that some differences in the cell surface of SE1045 and *E. coli* might cause the different binding affinity with 3-MPBA. To test this hypothesis, more experiment and discussion were conducted in the next section.

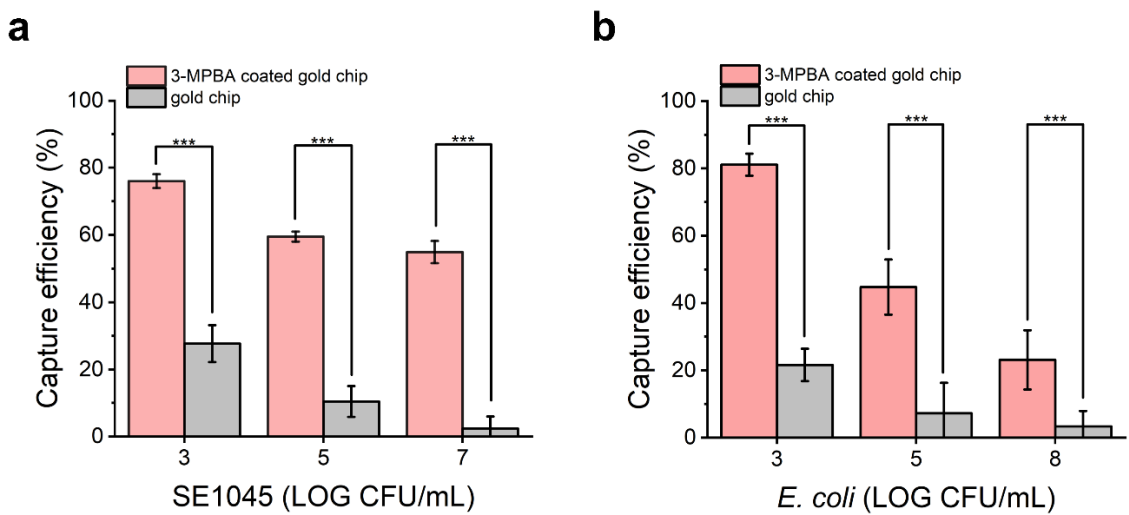

Figure 3. Capture efficiency of the 3-MPBA coated gold chip and bare gold chip. (a). The percentage of SE1045 captured by the gold chip with 3-MPBA coating (pink) and without 3-MPBA coating (gray) at $10^3$, $10^5$ and $10^7$ CFU/mL. (b) The percentage of *E. coli* captured by the gold chip with 3-MPBA coating (pink) and without 3-MPBA coating (gray) at $10^3$, $10^5$ and $10^8$ CFU/mL. Statistically significant differences between the capture efficiency of 3-MPBA coated gold chip and bare gold chip was determined by paired t-test. Asterisks illustrated the significance levels of *p <= 0.05, **p <= 0.01, ***p <= 0.001. Data were represented as mean ± SD (n = 3).

### 3.3 Quantification performance

To evaluate the quantification capability and understand the limit of detection of our developed method, a series of log bacteria concentrations ranging from $10^0$ to $10^8$ CFU/mL were implemented to sample preparation and optical analysis. Figure 4a exhibited the optical images of bacterial cells at different concentrations, which were processed and analyzed by ImageJ. Threshold value was set to cover every particle in the images with yellow area. As a result, the counted "black dot" was marked with green line (Figure 4b). The good linear relationship was found between the counting result and the corresponding theoretical bacterial concentration with an $R^2$ value of 0.9581 in terms of SE1045, and 0.9836 in terms of *E. coli* (Figure 4e and 4f). Due to the higher capture efficiency of 3-MPBA coated gold chip towards SE1045 compared to that of *E. coli* at high concentration, the single SE1045 cell was hard to be identified since the quantity of bacterial cells was too large to clump with each other. Consequently, the maximum quantifiable concentration was $10^7$ CFU/mL for SE1045, while it extended to $10^8$ CFU/mL for *E. coli*. Nevertheless, the limit of detection (LOD) of this assay was consistent for both SE1045 and *E. coli*, achieving as low as $10^3$ CFU/mL which shown significant difference ($p < 0.05$) with the lower concentrations. Literally, the "black dot" cannot be observed on the images at $10^0$ CFU/mL, but it appeared and the amount of it was consistent with the amount of $10^1$ and $10^2$ CFU/mL. The constant number of "black dots" was the background noise inherent of our method which might contributed from dust in the air and crystals from ammonia bicarbonate solution. Overall, these results indicated our method was reliable and sensitive to quantify bacterial cells across a broad range of concentrations.

### 3.4 Improvement of capture efficacy of *E. coli* using heat treatment

As previously reported (Ebrahimi et al., 2018; Katsui et al., 1982; Mcdaniel et al., 2019; Mitsuzawa et al., 2006), the heat treatment induced changes in bacterial cells including increasing hydrophobic property, partial loss of the outer membrane contributing to disorganization of the structure of outer membrane, and decreasing negative surface charges to some extent. The decreased charges corresponded to a depolarization, causing the enhanced hydrophobicity identically. Based on the principle of 3-MPBA binding to glycans on the cell surface and hydrophobicity of 3-MPBA, we hypothesized that the number of captured cells would increase after heat treatment because the hydrophobic molecules interacted easier with the hydrophobic one and peptidoglycan exposed more without the outer membrane, resulting in the higher sensitivity and lower LOD of our assay. To test this assumption, the bacterial cells was heated at 72 °C for 10 min prior to incubating with 3-MPBA coated gold chip. As evidenced in Figure 5a and 5b, the count of *E. coli* cells with heat treatment captured by the 3-MPBA coated gold chip was significantly higher than the cells without heat treatment. It was worth to note that the *E. coli* cells sedimented to the bottom after heat treatment rather than suspension, which might be attributed from the increased hydrophobicity making it less stable in an aqueous medium. As a consequence, the LOD of our method went down 1 LOG CFU/mL for heated *E. coli* cells and the great linear correlation between bacterial concentrations and counting results occurred in the range of $10^2$ to $10^7$ CFU/mL ($R^2$ = 0.9349) (Figure 5c). As for SE1045, there was no change between the number of captured cells irrespective of heat treatment (Figure 5d and 5e), even the suspending behavior of the bacterial cells in medium. It was reasonable that the quantification capability remained unchanged, and the detectable range of concentrations was the same as that of the untreated cells (Figure 5f). The images and results obtained from inactive cells further proved our method was effective to detect bacterial cells independent of their viability.

The distinct responses observed between SE1045 and *E. coli* cells under the same heat treatment strongly suggested the presence of differences in the surface between the two species, such as variations in charge abundance, hydrophobicity degree, or glycan content. Such differences strongly coordinated with the differences in the results of capture efficiency. Furthermore, *E. coli* cells was more heat-sensitive than SE1045, consisting with the findings from other research (Suhalim et al., 2023). To validate the difference in surface characteristics, we investigated the surface charge of bacterial cells before and after heat treatment using the Zeta potential measurement (Halder et al., 2015). Generally, the net surface charge of most bacteria was negative (Figure S7). The lower of the Zeta potential of *E. coli* (- 56.3 ± 1 mV) compared to the SE1045 (- 9.8 ± 1.5 mV) represented the higher abundance of negative charges with *E. coli*. Following heat treatment, the Zeta potential of *E. coli* cells (- 38.3 ± 0.2 mV) increased significantly, indicating the loss of negative charge and an increase in hydrophobicity, resulting in the higher binding affinity with 3-MPBA. In contrast, heat treatment had no discernible impact on the Zeta potential of SE1045, which corresponded with the observed consistency in capture efficiency and LOD. Moreover, we speculated that the decreased surface negative charge and enhanced hydrophobic property of bacterial cells may not the only factors contributing to the sedimentation because the SE1045 cells always kept suspending in medium even though the Zeta potential of the SE1045 cells was lower than the *E. coli* cells. Other variables might play a role of the sedimentation, such as the density of cells, leakage of the cellular content, and environmental effects caused by heating, which should be confirmed with further experiment (Mcdaniel et al., 2019; Mitsuzawa et al., 2006).

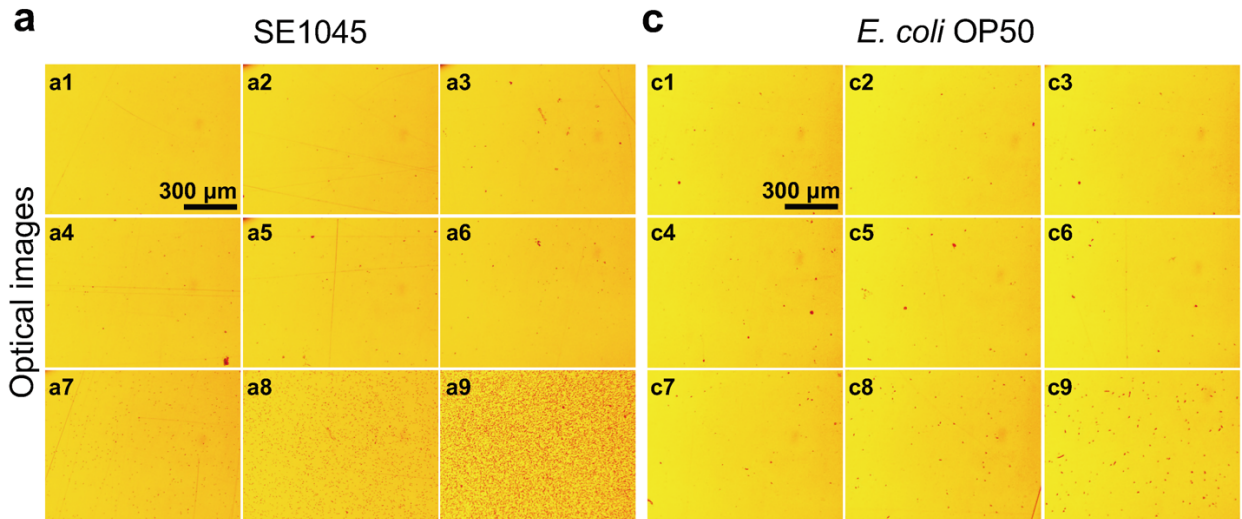
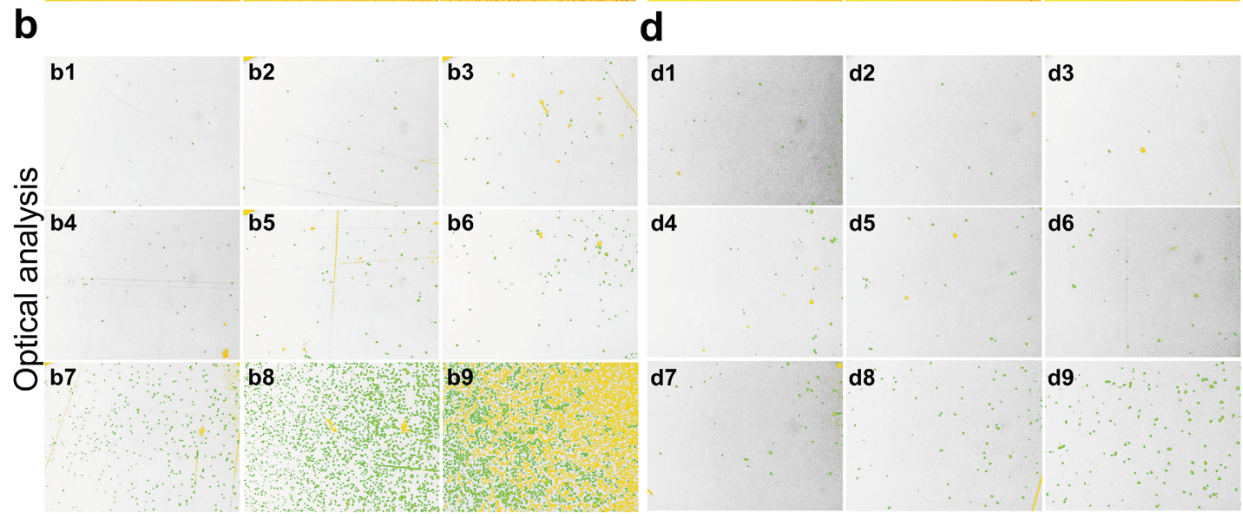
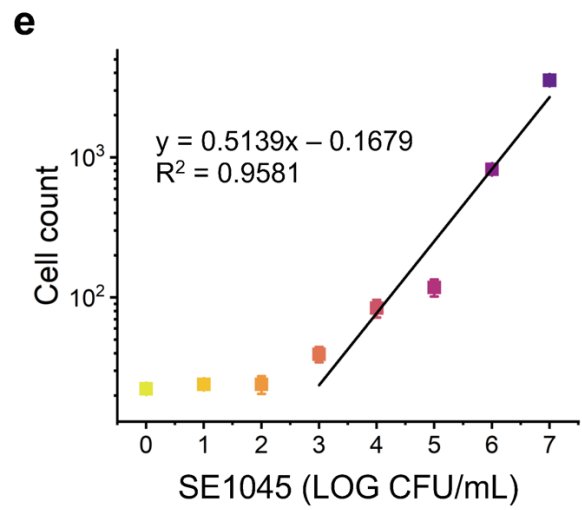
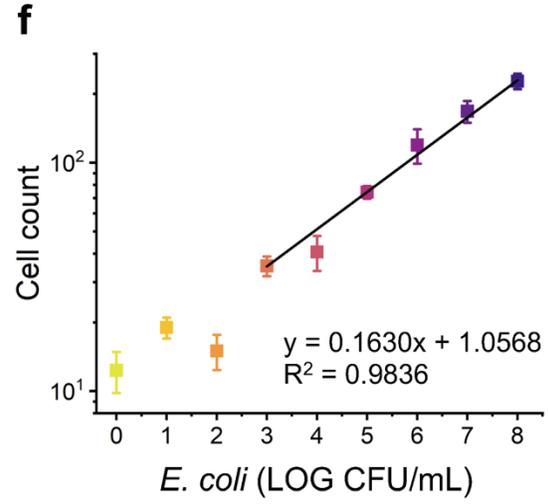

Figure 4. Optical images and optical analyzed images of bacterial cells at different concentrations. (a) Optical images of SE1045 cells on 3-MPBA coated gold chip at different concentrations (a1) $10^0$ CFU/mL (a2) $10^1$ CFU/mL (a3) $10^2$ CFU/mL (a4) $10^3$ CFU/mL (a5) $10^4$ CFU/mL (a6) $10^5$ CFU/mL (a7) $10^6$ CFU/mL (a8) $10^7$ CFU/mL (a9) $10^8$ CFU/mL; (b) Optical analysis images obtained after the ImageJ processing with the same concentrations mentioned in (a); (c) Optical images of *E. coli* OP50 cells on 3-MPBA coated gold chip at different concentrations (c1) $10^0$ CFU/mL (c2) $10^1$ CFU/mL (c3) $10^2$ CFU/mL (c4) $10^3$ CFU/mL (c5) $10^4$ CFU/mL (c6) $10^5$ CFU/mL (c7) $10^6$ CFU/mL (c8) $10^7$ CFU/mL (c9) $10^8$ CFU/mL; (d) Optical analysis images obtained after the ImageJ processing with the same concentrations mentioned in (c); (e) Correlation between SE1045 concentrations and the cell counts calculated from images; (f) Correlation between *E. coli* OP50 concentrations and the cell counts calculated from images. Data are represented as mean ± SD (n = 4). Scale bar represented 300 μm length.

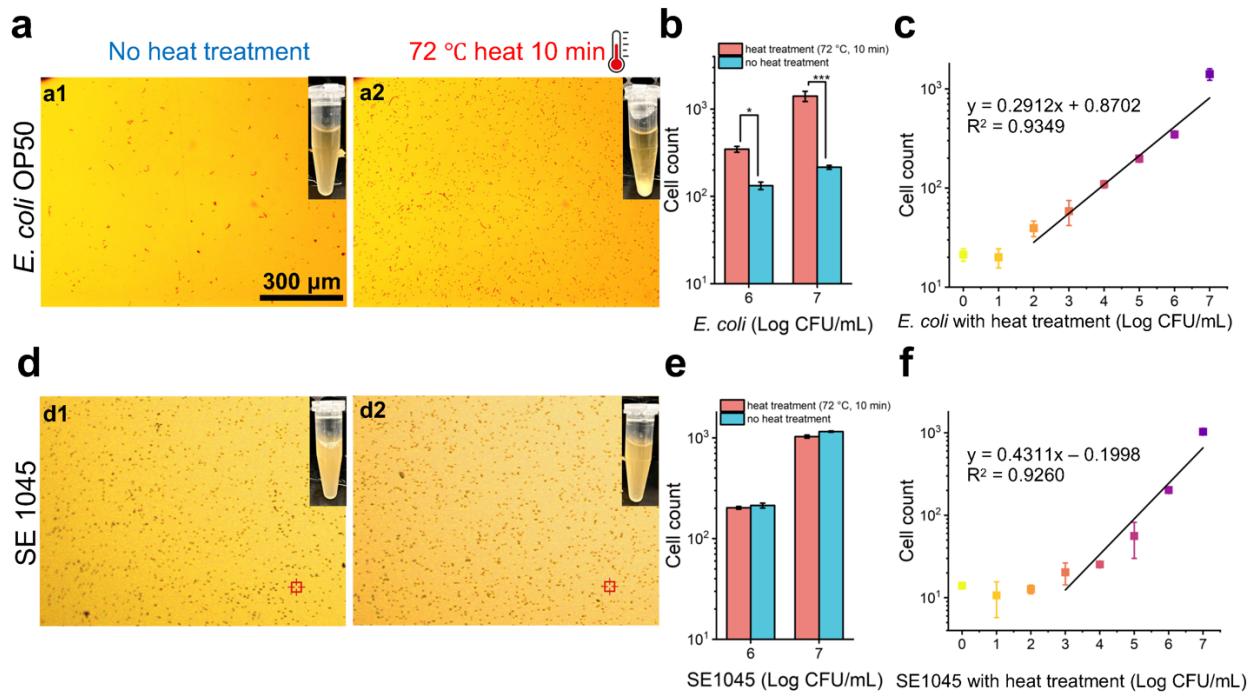

Figure 5. Comparison of optical images and cells counts of bacterial cells with/without heat treatment captured by 3-MPBA coated gold chip. (a) Optical images of *E. coli* cells without (a1) and with (a2) heat treatment on 3-MPBA coated gold chip. (b) Comparison of cell counts of *E. coli* with/without heat treatment at $10^6$ CFU/mL and $10^7$ CFU/mL (*$p <= 0.05$ **$p <= 0.01$ ***$p <= 0.001$). (c) Correlation between concentrations of heated *E. coli* OP50 and the cell counts calculated from images. (d) Optical images of SE1045 cells without (d1) and with (d2) heat treatment on 3-MPBA coated gold chip. (e) Comparison of cell counts of SE1045 with/without heat treatment at $10^6$ CFU/mL and $10^7$ CFU/mL. (f) Correlation between concentrations of heated SE1045 and the cell counts calculated from images. Data are represented as mean ± SD (n = 4). Scale bar represented 300 μm length.

## 3.5 Swab recovery and detection of bacterial cells seeded on stainless-steel surface

To investigate the feasibility of detecting and quantifying bacterial cells on surfaces, we implemented swab sampling prior to optical detection to collect SE1045 cells of $10^4$ CFU which was distributed evenly on the 10 × 10 cm area of stainless-steel surface. Since the foam swabs has been described to show the high recovery and release capacity due to the open structure (Jansson et al., 2020), it was employed to implement swabbing collection. SE1045 cells were collected by the water-premoistened foam swab, then released in ammonia bicarbonate through vigorous vortex, and incubated with 3-MPBA coated gold chip subsequently. To evaluate the recovery rate of bacterial cells, several controls were set to do comparison and calculation, including in the positive control: ammonia bicarbonate with $10^4$ CFU SE1045 cells; negative control: ammonia bicarbonate without SE1045 cells; and negative control for recovery from surface: swabbing the stainless-steel surface without SE1045 cells. More particles can be observed on the images of the samples recovered from stainless-steel surface (Figure 6c) in comparison of positive and negative controls (Figure 6a). It indicated plenty of interference particles produced which might originated from the fabric residues of swab caused by friction during swabbing. Apparently, some particles were non-round shape, in darker or lighter color, and in a larger size, which were easily to be differentiated with bacterial cells and excluded by the ImageJ with specified particle parameters (See section 2.3) (Figure 6d). However, some interference particles were still counted due to the high similarity with bacterial cells and limited ability of ImageJ to tell the subtle differences, leading to the number of particles of negative control recovered from surface was much higher than the plain negative control (Table 1). Thus, we assumed the interference level was constant in every image and subtracted by the number of particles of actual sample to calculate the accurate count of bacterial cells. As the result, the recovery rate reached up to $70.6 \pm 6.2$ %, with the entire procedure taking approximately 2 hours to complete. Due to the small size of the gold chip we used, the cost per sample was remarkably low, estimated at $ 2 including the foam swab, gold chip and chemical reagents.

This application illustrated the feasibility and cost-effectiveness of integrating optical detection with swab recovery for the purpose of detecting bacterial cell loading on stainless-steel surface in 2 hrs. However, it is the initial attempt, and more conditions of swabbing should be optimized and explored in the future study to achieve better recovery rates and minimize the interference produced from the swabbing process. According to the current studies, the physical force of swabbing, wetting solution to moist swab, drying time of bacterial solution remaining on the surface, releasing solution and approaches to assist releasing shown great impact on swabbing outcomes, which should be considered (Jansson et al., 2020; Jones et al., 2020; Landers et al.,

2010). Furthermore, while ImageJ proved effective in cell counting, it faced the limitation in differentiating the particles that looks similar with the bacterial cells. It will become the challenge when it comes to be applied in the real world, especially facing the complex food matrices. Fast developing artificial intelligence and machine learning models might provide the solution to distinguish the bacterial cells and other particles, even categorize the different food residues (Deng et al., 2021; Yan et al., 2021; Yang et al., 2021). Despite the total time of our assay was notably shorter than traditional plate counting, unfortunately, it still longer than the ATP swabbing kit. To save more time and enhance the user-friendliness, more efforts should be done towards simplifying of procedure without compromising accuracy as much as possible.

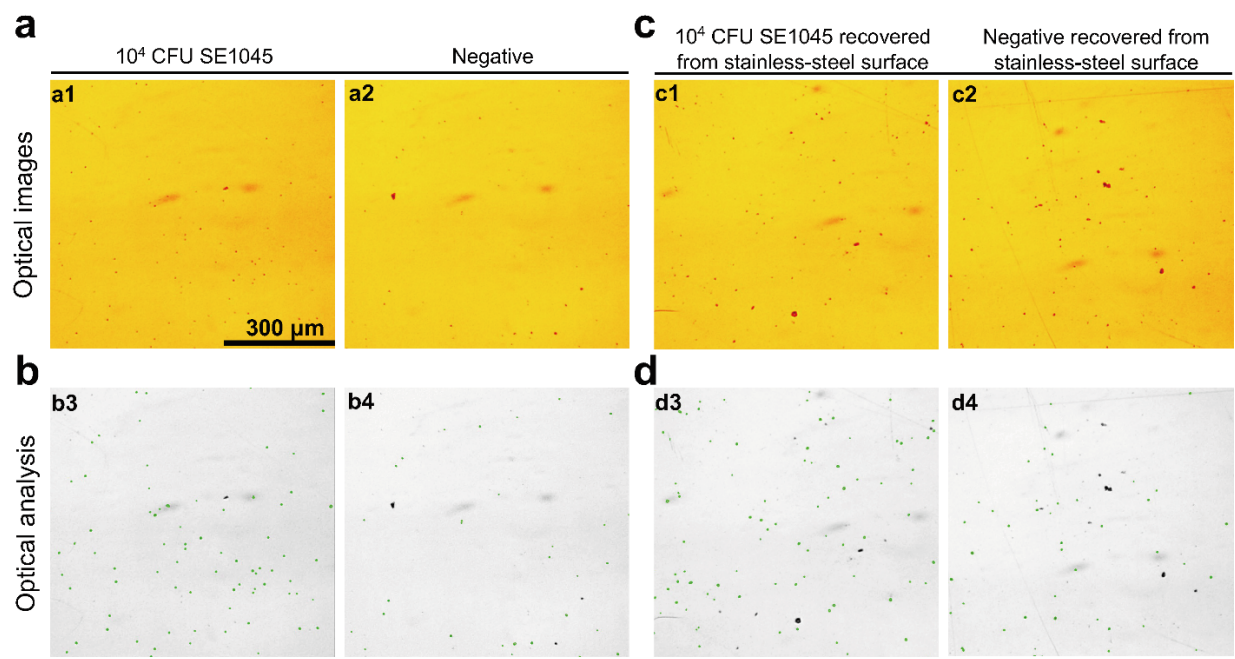

Figure 6. Optical detection of SE1045 cells recovered from stainless-steel surface. (a) Optical images of (a1) $10^4$ CFU SE1045 and (a2) negative control (ammonia bicarbonate without SE1045 cells) on 3-MPBA coated gold chip. (b) Optical analysis images processed from (a) by ImageJ. (c) Optical images of (c1) $10^4$ CFU SE1045 recovered from stainless-steel surface and (c2) negative

control (no SE1045 cells) recovered from stainless-steel surface on 3-MPBA coated gold chip. (d) Optical analysis images processed from (c) by ImageJ. Scale bar represented 300 μm length.

| Samples | Counts | Counts (Positive - Negative) | Recovery (%) |
|---|---|---|---|
| | Avg ± SD | Avg ± SD | Avg ± SD |
| $10^4$ CFU SE1045 | 84.5 ± 14.8 | 58.0 ± 14.5 | 70.6 ± 6.2 |
| Negative | 30.7 ± 5.2 | | |
| $10^4$ CFU SE1045 (recovered from stainless-steel surface) | 143.3 ± 33.6 | 40.7 ± 9.3 | |
| Negative (recovered from stainless-steel surface) | 102.7 ± 36.3 | | |

Table 1. Recovery of SE1045 cells loading on stainless-steel surface of 10 × 10 cm area. Data are represented as mean ± SD (n = 4).

**Conclusion**

In this study, the optical detection method was established based on the 3-MPBA coated gold chip for rapid bacterial detection with low magnification objective lens (10 ×) of general optical microscope. The basic steps included: 1. coating 3-MPBA on gold chip overnight; 2. washing away excess 3-MPBA; 3. incubation of bacterial cells with 3-MPBA coated gold chip for 1 hour; 4. wash away ammonia bicarbonate crystals and clumped bacterial cells. By optimizing those conditions, the single bacterial cells can be observed and counted. The capture efficiency of 3-MPBA coated gold chip was significantly higher than the uncoated gold chip regardless of SE1045 or *E. coli* cells. Moreover, our method exhibited good quantification capability with the limit of detection of $10^3$ CFU/mL for both SE1045 and *E. coli* cells. The detection limit of *E. coli* cells improved by 1 LOG CFU/mL through heat treatment. In the last, it was successful in combining with swab recovery to detect SE1045 cells of $10^4$ CFU loading on the stainless-steel surface with a 70 % recovery rate in 2 hrs. The total cost of material in this procedure was only $ 2 per sample and $ 250 for the optical microscope. Hence, our established method showed great potential to be applied to detect bacterial cells recovered from the surface in a rapid, effective, and low-cost way, applicable in limited resource settings. Future studies will focus on exploring the optimized conditions of swab recovery, methods to differentiate bacterial cells from other food residues in images, and simplifying the procedure to make it more user-friendly and practical.

**Author contributions**

Yuzhen Zhang: Methodology, Formal analysis, writing – original draft, Methodology. Zili Gao: Methodology. Lili He: Conceptualization, Supervision, Writing – review & editing.

**Acknowledgment**

We thank Prof. Lynne McLandsborough at the University of Massachusetts Amherst for generously donating SE1045 strains and Prof. Yeonhwa Park for *E. coli* OP50 strains. The first author Yuzhen Zhang was supported by the Tan fellowship.

destructive paper chromogenic array detection of multiplexed viable pathogens on food. Nat Food 2, 110–117. https://doi.org/10.1038/S43016-021-00229-5

Supporting information

# Optical detection for bacterial cells on stainless-steel surface with a low-magnification light microscope

**Authors:** Yuzhen Zhang[a], Zili Gao[a], and Lili He[*a]

**Affiliations:**

[a]Department of Food Science, University of Massachusetts, Amherst, MA 01003, USA.

[*] Co-corresponding authors:

Dr. Lili He, E-mail: lilihe@foodsci.umass.edu; Fax: +1 413 545 1262; Tel: +1 413 545 5847

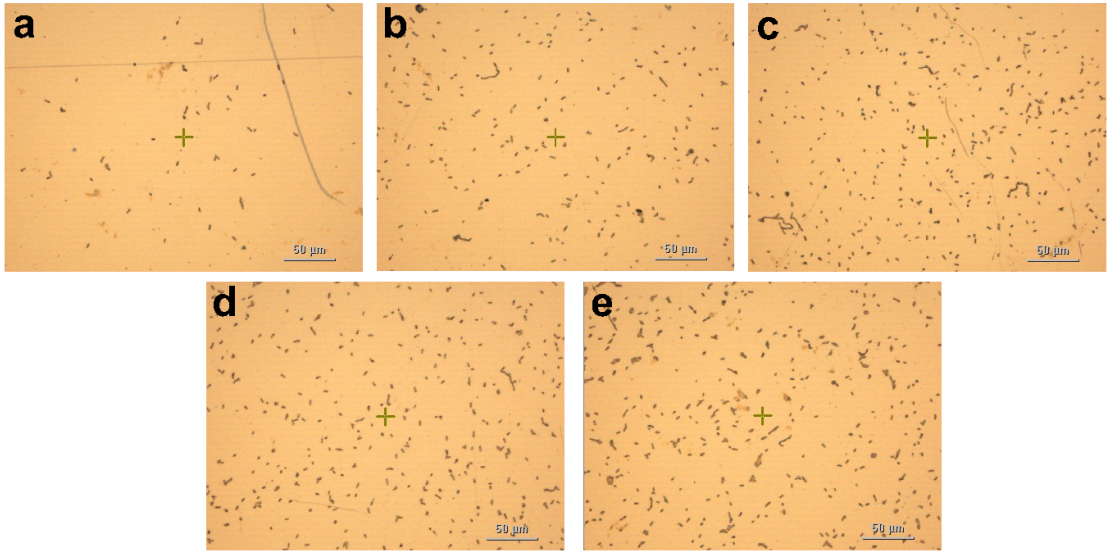

Figure S1. Optical images of *E. coli* cells at $10^8$ CFU/mL on gold chip coated with different concentrations of 3-MPBA. (a) 10 mM; (b) 20 mM; (c) 40 mM; (d) 80 mM; (e) 160 mM of 3-MPBA. These images were taken under 20 × objective lens. Scale bar represented 50 μm length.

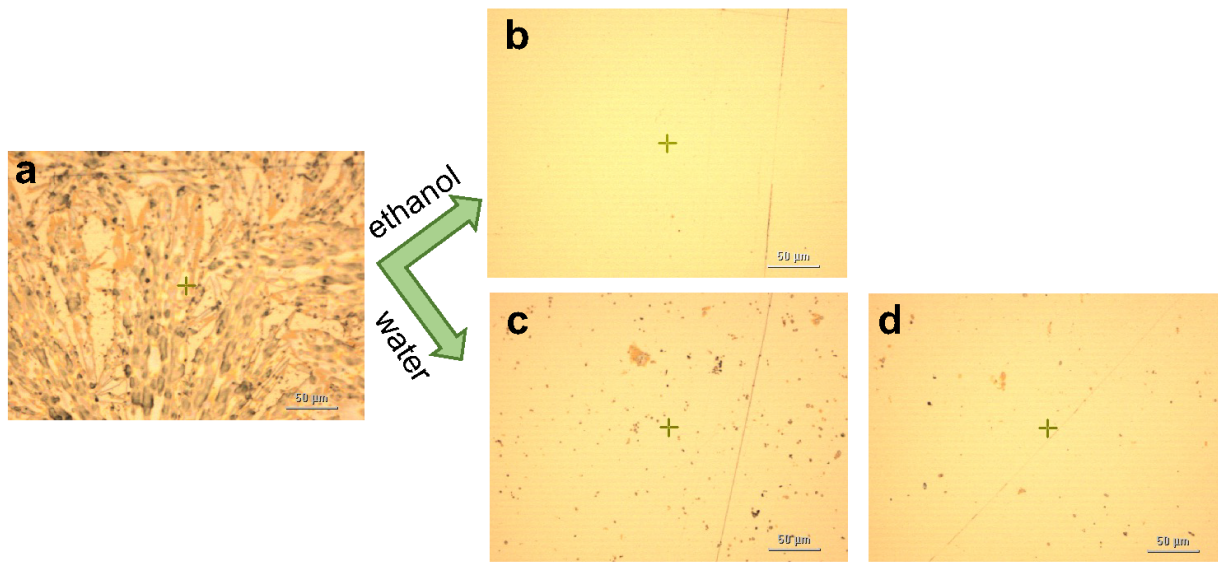

Figure S2. Comparison of optical images of 3-MPBA coated gold chip after washed by different solutions. (a) Gold chip coated with 3-MPBA overnight without washing. (b) 3-MPBA coated gold chip washed once with ethanol. (c) 3-MPBA coated gold chip washed once with water. (d) 3-MPBA coated gold chip washed twice with water. These images were taken under 20 × objective lens. Scale bar represented 50 μm length.

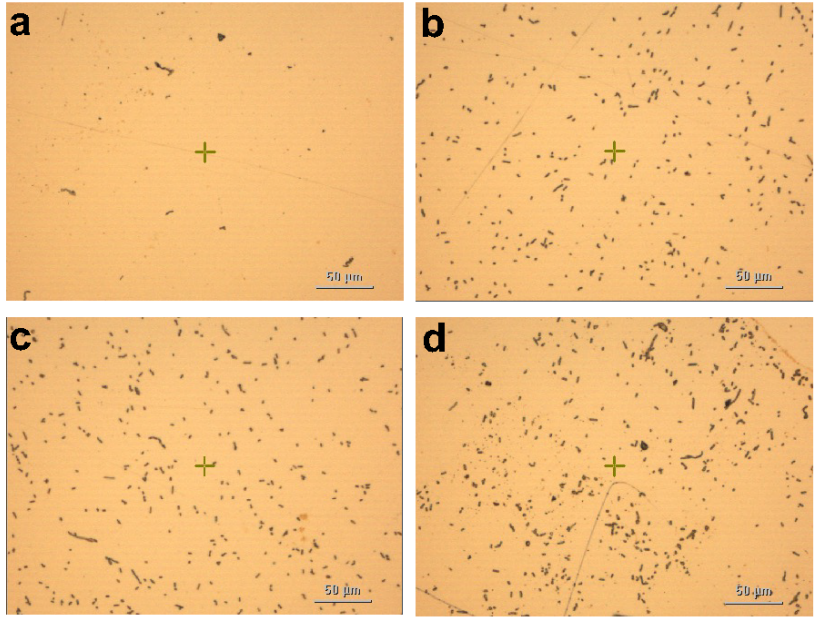

Figure S3. Optical images of different incubation time of *E. coli* cells at $10^8$ CFU/mL and 3-MPBA coated gold chip. (a) 30 minutes; (b) 60 minutes; (c) 90 minutes; (d) 120 minutes. These images were taken under 20 × objective lens. Scale bar represented 50 μm length.

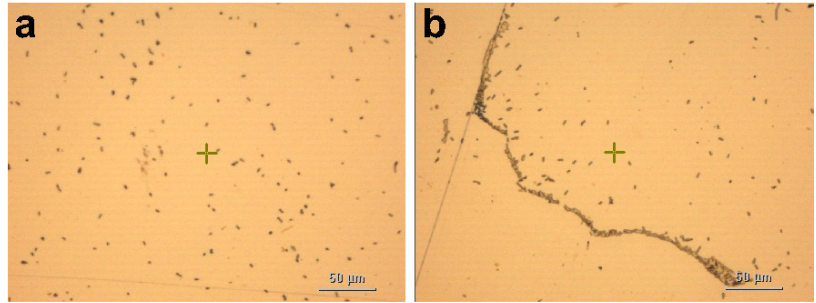

Figure S4. Comparison of optical images of 3-MPBA coated gold chip with wet/dry surface capturing *E. coli* cells at $10^8$ CFU/mL. (a) *E. coli* cells were captured with wet surface 3-MPBA coated gold chip. (b) *E. coli* cells were captured with dry surface 3-MPBA coated gold chip. These images were taken under 20 × objective lens. Scale bar represented 50 μm length.

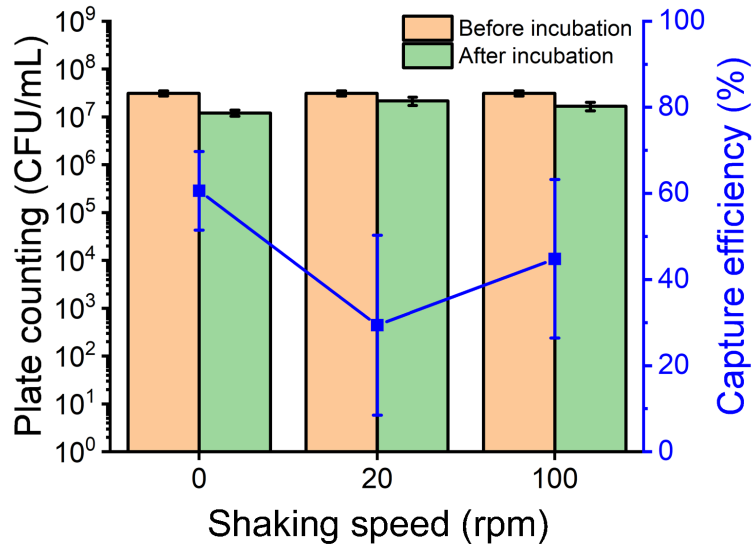

Figure S5. Bacterial capture efficiency of 3-MPBA coated gold chip with different shaking speeds during incubation. The number of bacterial cells before (orange bar) and after (green bar) incubating with 3-MPBA coated gold chip was determined by plate counting. The calculation of capture efficiency based on the results of plate counting. Data are represented as mean ± SD (n = 3).

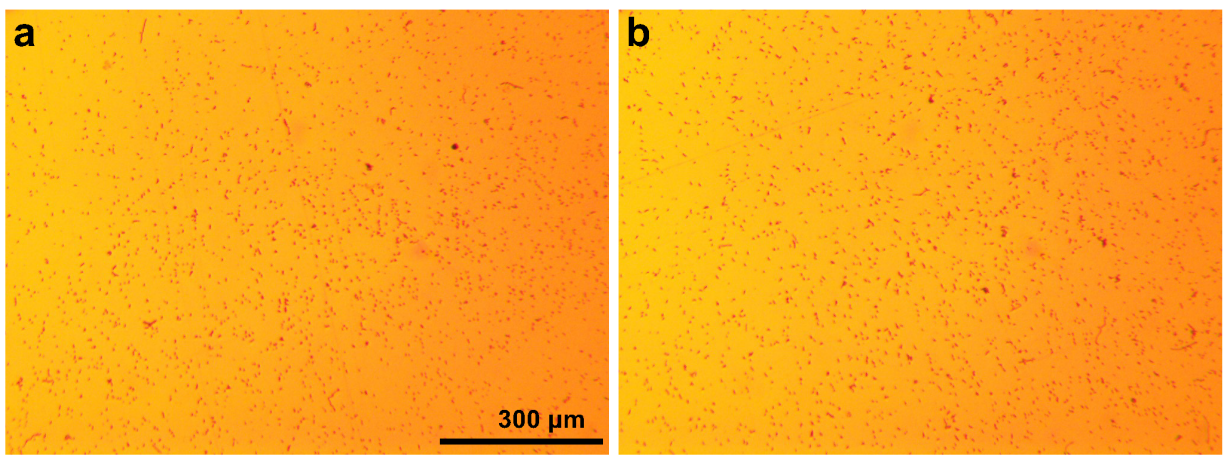

Figure S6. Comparison of optical images of washing times after incubation of SE1045 at $10^8$ CFU/mL with 3-MPBA coated gold chip. (a) Washing twice with water. (b) Washing three times with water. These images were taken by 10 × objective lens. Scale bar represented 300 μm length.

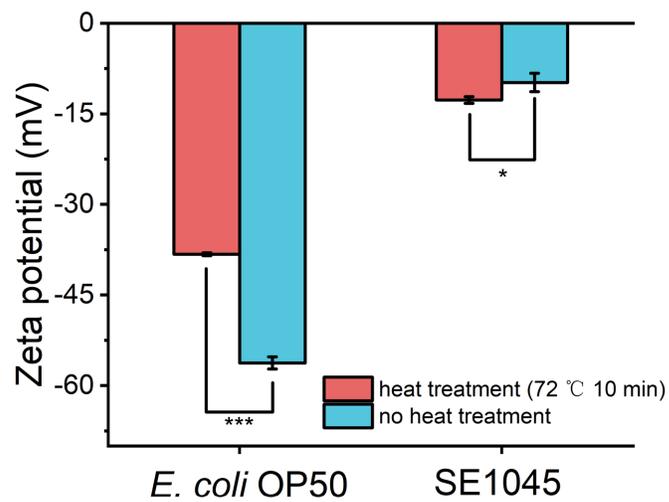

Figure S7. Zeta potential of *E. coli* and SE1045 cells with and without heat treatment. Data are represented as mean ± SD (n = 3).